\documentclass[twocolumn,showpacs,preprintnumbers,amsmath,amssymb]{revtex4}

\usepackage{dcolumn}
\usepackage{bm}
\def\ee{\end{equation}}                       
\def\be{\begin{equation}}
\def\ba{\begin{eqnarray}} 
\def\ea{\end{eqnarray}}

\begin{document}

\title{Equivalence of  nonadditive entropies and  nonadditive energies
in long range interacting systems under macroscopic equilibrium}
\author{Ramandeep S. Johal}
\email{rjohal@jla.vsnl.net.in}
 \affiliation{Physics Department, Lyallpur    Khalsa    College,
Jalandhar-144001, India.}
\author{Antoni Planes}
\email{toni@ecm.ub.es}
\author{Eduard Vives}
\email{eduard@ecm.ub.es}
\affiliation{ Departament d'Estructura i Constituents de la Mat\`eria,
Universitat  de Barcelona  \\ Diagonal  647, Facultat  de F\'{\i}sica,
08028 Barcelona, Catalonia}
\date{}

\begin{abstract}
We discuss  that the thermodynamics of composite  systems with non-additive
entropies  and additive  energies can  be equivalently  derived considering
additive entropies  and non-additive energies.  The general discussion
is illustrated by a particular example.
\end{abstract}


\maketitle
%
\section{Introduction}

Additivity of quantities  such as entropy and energy,  is an essential
premise  of  classical  thermodynamics  and statistical  mechanics  of
systems  with short  range interactions  \cite{Callen1985}.  Arguments
about  positivity  conditions  of  specific  heat are  based  on  this
additivity  postulate.  On the  other hand,  systems where  long range
interactions  are  relevant   or  with  statistical  correlations  are
basically  nonadditive in  the  energy and/or  entropy.   Due to  this
feature, such systems  display unusual properties, like inequivalence
of  the different statistical  mechanics ensembles,  negative specific
heat in microcanonical ensemble and possible temperature discontinuity
at  first  order   transitions  \cite{Dauxois2002}.   For  nonadditive
systems, the non-concave entropy has  been suggested to be the correct
entropy of the  system, with the result that the  specific heat may be
negative \cite{Antonov1962, Lynden1968, Lynden1999}.

Similarly,  the  notion  of  macroscopic  thermal  equilibrium,  which
introduces the concept of temperature through the zeroth law, is based
on additivity  of entropy  and energy.  On  the other hand,  for large
systems with long range interactions or even finite systems with short
range interactions, the concept of equal temperatures over subsystems,
does not hold at thermal  equilibrium (defined as the maximum of total
entropy).   Clearly, if the additivity  postulate is  relaxed for  either
entropy or energy, we have a breakdown of the standard zeroth law.

This paper  is divided into  three sections. In section  \ref{sec2}, we
present  how within  a statistical  mechanical framework  of correlated
systems, one can  interpret the fact  that the  standard Boltzmann
entropy functional renders non-additive macroscopic entropies and thus
a different  entropy functional may  lead to macroscopic  entropy with
the additive  character.  This  analysis is done  keeping in  mind the
examples  recently  presented   by  C.Tsallis  \cite{Tsallis2004} which
justify that  for a particular kind of correlations,  Tsallis entropy
functional does behave additively.   In section \ref{sec3}, we analyze
from a  macroscopic thermodynamics point of view,  how the equilibirum
state (zero principle) of a composite system with non-additive entropy
and  additive energy  functions can  be equivalently  described  by an
additive entropy but a  non-additive energy.  In section \ref{sec4}, we
present a  particular example of this equivalence.  Finally in section
\ref{sec5}, we  conclude that the description of  the equilibrium state
of  a system  with a  non-Boltzmann entropy  with  additive properties
requires  the  use   of  a  energy  functional  which   will  lead  to
non-additive energies.

\section{Statistical framework}
\label{sec2}
Let  us  consider   two  systems  $A$  and  $B$   with  energy  levels
$\varepsilon^A_i  \in  \Omega^A$  and $\varepsilon^B_j  \in  \Omega^B$
whose  probability  distribution  functions  are $p^A_i$  and  $p^B_j$.
Macroscopically,  thermodynamics  of  each  of  these  systems  can  be
developed,  using the  standard  averages and Boltzmann entropy. This is:
\begin{equation}
E^{A(B)}   =  \langle   \varepsilon^{A(B)}  \rangle   =   \sum_{k  \in
\Omega^{A(B)}} p^{A(B)}_k \varepsilon^{A(B)}_k,
\label{standardav}
\end{equation}
and
\begin{equation}
S^{A(B)} = \sum_k p^{A(B)}_k \ln p^{A(B)}_k.
\end{equation}
Now let us consider the composite system $A+B$ with energy states in 
$\Omega^A\times \Omega^B$ and assume that its energy levels are given by
\begin{equation}
\varepsilon^{A+B}_{ij}=\varepsilon^{A}_{i}+\varepsilon^{B}_{j},
\end{equation}
but the probability distribution of the composite system exhibits some 
statistical correlations, so that 
\begin{equation}
p^{A+B}_{ij} \neq p^A_i p^B_j.
\end{equation} 
It is clear that for the composite system the mean energy satisfies
\begin{equation}
E^{A+B}=\sum_{ij}            p^{A+B}_{ij}            \left           (
\varepsilon^{A}_{i}+\varepsilon^{B}_{j}\right )=E^A+E^B,
\end{equation}
where we have used the fact  that $p^A_i$ and $p^B_j$ are the marginal
probabilities of $p^{A+B}_{ij}$. In  contrast, it is straightforward to
see that the entropy of the composite system is non-additive, i.e.
\begin{equation}
S^{A+B}\neq S^A+S^B.
\end{equation}
Let us  now suppose that  we can introduce a  non-Boltzmann entropy
functional of  the probability distribution $\tilde  S(p_i)$ such that
it satisfies additivity for the composite system, i.e.
\begin{equation}
\tilde S^{A+B}= \tilde S^A+ \tilde S^B.
\end{equation}
This idea is  based on a recent paper  by C. Tsallis \cite{Tsallis2004},
where  for  systems with  specific  correlations,  the Tsallis  entropy
functional is shown to  satisfy the additivity relation. The aim of
the  present paper  is to  show, using  thermodynamic  arguments (zero
principle),  that  if  one  describes  the composite  system  with  an
additive $\tilde  S$, then one  should modify the average  energies so
that they become non-additive
\begin{equation}
\tilde E^{A+B} \neq \tilde E^A+ \tilde E^B.
\end{equation}
This means that  the ``averages'' cannot be the  standard ones defined
in  Eq. (\ref{standardav}). For  instance, in  the case  of  systems with
specific  correlations  for  which  Tsallis entropy  is  additive,  we
conjecture that  the new  averages would correspond  to escort-averages
which will render non-additive average energies.
%
\section{Thermodynamic framework}
\label{sec3}
%
We  consider two  thermodynamic systems  $A$ and  $B$  with individual
equations  of state  $S^A=S^A(E^A)$ and  $S^B=S^B(E^B)$ for  which the
energy of the composite system ($A+B$) is additive
\be
E^{A+B} = E^{A}+ E^B,
\ee
but the entropy of the  system ($A+B$) is non-additive and can
be written as:
\be
S^{A+B} = S^A + S^B + F(S^A, S^B).
\ee
When  the two  systems $A$  and $B$  are in  ``thermal  contact'', the
macroscopic   state  of   the  composite   system  is   obtained  upon
maximization of  the total entropy  under the constraint  of constant
energy. Therefore
\be
dE^{A+B}= 0 \Rightarrow dE^A = - dE^B.
\label{eab}
\ee
The maximum entropy condition ($dS^{A+B}=0$) yields:
\ba
0 & = & \frac{dS^A}{dE^A} dE^A + \frac{dS^B}{dE^B} dE^B+ 
\nonumber \\
&+& \frac{\partial F}{\partial S^A} \frac{dS^A}{dE^A} dE^A 
+ \frac{\partial F}{\partial S^B} \frac{dS^B}{dE^B} dE^B.
\ea
Using (\ref{eab}) one gets the zero principle (or equilibrium) 
condition for this system:
\be
\left ( 1+ \frac{\partial F}{\partial S^A} \right ) \frac{dS^A}{dE^A}= 
\left ( 1+ \frac{\partial F}{\partial S^B} \right ) \frac{dS^B}{dE^B}.
\label{zero1}
\ee
Note that if  one considers the usual definition  of the inverse local
temperature    of    each    system   $\alpha$    ($\alpha=A,B$)    as
$\frac{1}{T^\alpha}    =    \frac{dS^\alpha}{dE^\alpha}$,   the    two
temperatures are not equal in equilibrium but the zero principle still
fixes  the ratio  between  them,  which will  depend  on the  function
$F(S^A, S^B)$.

Let us now assume that the  same two systems can be described assuming
that the entropy  is additive, while the non-additivity  is only on the
energy, through a function $G(E^A, E^B)$ so that:
\be
{\cal E}^{A+B}=E^A+E^B+G(E^A,E^B),
\ee
\be
{\cal S}^{A+B}=S^A+S^B.
\ee
The question is, whether such  a description of the joint system will
lead to  the same zero  principle. Imposing again maximization  of the
composite entropy (${\cal S}^{A+B}$)  under the  constraint of constant 
composite energy (${\cal E}^{A+B}$), we obtain:
\be
\left ( 1+ \frac{\partial G}{\partial E^B} \right ) \frac{dS^A}{dE^A}= 
\left ( 1+ \frac{\partial G}{\partial E^A} \right ) \frac{dS^B}{dE^B}.
\label{zero2}
\ee
Now, if we demand that the same zero
principle given by (\ref{zero1})  is satisfied in both descriptions, then
in general, the following conditons should hold:
\ba
\left ( 1+ \frac{\partial G}{\partial E^B} \right ) &=& I(E^A, E^B)
\left ( 1+ \frac{\partial F}{\partial S^A} \right ), \label{k1} \\
\left ( 1+ \frac{\partial G}{\partial E^A} \right ) &=& I(E^A, E^B)
\left ( 1+ \frac{\partial F}{\partial S^B} \right ), \label{k2}
\ea
where $ I(E^A, E^B)$ is an unknown differentiable function of its arguments.
Clearly, the above conditions imply that Eq. (\ref{zero2}) is the same
condition as Eq. (\ref{zero1}).
In the following, we fix the function $I$ to be a constant equal to unity,
by noting the fact that for the case of standard thermodynamics, we have 
$F=0$, implying that $G=0$. With this choice, Eqs. (\ref{k1}) and (\ref{k2})
lead to two independent conditions:
\be
\frac{\partial G}{\partial E^B}= \frac{\partial F}{\partial S^A},
\label{g1}
\ee
and
\be
\frac{\partial G}{\partial E^A}= \frac{\partial F}{\partial S^B}.
\label{g2}
\ee
Are  these two conditions  enough to  find the  function $G$,  given a
certain  function $F$ ? The  answer is  yes, provided  that $dG$  is a
exact differential. This means that the second cross derivatives of $G$
must  be equal.   This renders  a condition  on the  original function
$F(S^A, S^B)$:
\be
\frac{\partial^2 F}{\partial {S^A}^2} \frac{dS^A}{dE^A} = 
\frac{\partial^2 F}{\partial {S^B}^2} \frac{dS^B}{dE^B}.
\label{condition}
\ee
If this condition holds, $G$ can be obtained by integration of the Eqs. 
(\ref{g1}) and (\ref{g2}). The formal result is
\ba
G(E^A, E^B) & = & G(E^A_0, E^B_0) + \nonumber \\
&+& \int_{E^A_0}^{E^A} \frac{\partial F}{\partial S^B} dE^A + 
\int_{E^B_0}^{E^B} \frac{\partial F}{\partial S^A} dE^B, 
\label{solG}
\ea
which  will be  independent of  the path  chosen for  integration. In
particular, note that for a  composite system for which the entropy is
$S^{A+B}=S^A + S^B + \lambda S^A S^B$  \cite{Tsallis1988}, the  condition
(\ref{condition}) is satisfied independently  of the state equations of
the  two subsystems. (The state equations for individual systems are required
for integrations in Eq.(\ref{solG}).)  Thus,  there  exists a  function $G$  
that will allow  a description  of the  same  thermal contact in terms of an 
additive entropy and a non-additive energy.
%
\section{Particular example}
\label{sec4}
In this section,  we take a specific model  for a non-additive entropy
and use the  general treatment presented so far  in order to translate
it to the corresponding non-additive energy framework.

Let us  assume a  general polynomial function  $F(S^A,S^B)$, including
terms up to second order
\ba
F(S^A, S^B) & = & a_A S^A+ a_B S^B+ \nonumber \\
&+ & b_{AB}S^A S^B + b_A (S^A)^2 + b_B (S^B)^2,
\ea
and the following equations of state for the individual systems:
\be
S^{\alpha}(E^\alpha) = c_{\alpha} E^\alpha + d_\alpha.
\ee
Note that from the  definition of $F$, the condition (\ref{condition})
is inmediately fullfilled if we require
\be
b_A c_A = b_B c_B \equiv K.
\ee
To determine $G$, we shall integrate the partial derivatives:
\be
\frac{\partial G}{\partial E^A}= a_B+ b_{AB}S^A(E^A) + 2 b_B S^B(E^B),
\ee
\be
\frac{\partial G}{\partial E^B}= a_A+ b_{AB}S^B(E^B) + 2 b_A S^A(E^A).
\ee
Integration yields:
\ba
G(E^A, E^B) & = & (h_A -1) E^A + (h_B-1) E^B + \nonumber \\
&+& J_{AB} E^A E^B + \frac{J_A}{2} (E^A)^2 + \frac{J_B}{2} (E^B)^2,
\nonumber \\
\ea
where the constants $h_A$, $h_B$, $J_{AB}$, $J_A$ and $J_B$ are given by
\be
h_A = 1 + a_B + b_{AB} d_A + 2 b_B d_B + \sigma_A, 
\ee
\be
h_B = 1 + a_A + b_{AB} d_B + 2 b_A d_A + \sigma_B, 
\ee
\be
J_{AB}= 2 K,
\ee
\be
J_A = \frac{1}{2} K \frac{b_{AB}}{b_A}, 
\ee
\be
J_B = \frac{1}{2} K \frac{b_{AB}}{b_B}, 
\ee
where $\sigma_A$ and $\sigma_B$ are integration constants. 
We note that the energy of the composite system is of the form:
\ba
E^{A+B} & =&  h_A E^A + h_B E^B + \nonumber \\
&+& J_{AB} E^A E^B + \frac{J_A}{2} (E^A)^2 + \frac{J_B}{2} (E^B)^2
\ea
which corresponds to the energy of two coupled paramagnets with long
range interactions. This  model has been recently studied in depth by
J. Oppenheim \cite{Oppenheim2003}. It is plausible that such a nonadditive
behaviour of the macroscopic energy should arise from a non-standard 
averaging scheme of the microscopic hamiltonian, different from Eq. 
(\ref{standardav}).
%
%
\section{Summary and conclusions}
\label{sec5}
In this paper, we have considered the problem of thermal contact between two
thermodynamic systems, in order to discuss the interplay between non-additive
entropies and non-additive energies to describe the same equilibrium state
of the composite system. We show that an equilibrium condition 
between  the two systems,
described by set of quantities ($E^A$, $S^A$) and  ($E^B$, $S^B$), can be
realized in two equivalent ways: treating the entropy of the total system
as non-additive, but with the constraint in the form of additive total energy,
or alternately, treating the entropy as additive but the constraint in the
form of nonadditive total energy. The thermodynamic analysis becomes simplified
by assuming that the total entropy is a function of the entropies of the 
subsystems $A$ and $B$ only. Similarly, the total energy is completely determined
from the system energies. Thus we get a reinterpretation of the nonadditive
entropy rule in terms of a specific nonadditive energy rule.
We illustrate this correspondence by taking a specific example and obtain
a composite energy which represents the energy of two paramagnets interacting
via long range interactions. As a special case, we also observe that Tsallis
type nonadditivity \cite{Tsallis1988} corresponds to the case when the 
individual paramagnetic systems experience long range interactions only 
within each system, but not between themselves.

\section{Acknowledgements}
This  work has received financial support from CICyT (Spain),
project MAT2004-1291 and DURSI (Catalonia), project
2001SGR00066. RSJ thanks A.K. Rajagopal for
useful discussions.

\end{document}